\documentclass[
aps,
preprint,
floatfix,
amsmath,
amssymb
]{revtex4}

\usepackage{times}

\usepackage{graphicx,epsf}

\def\bkfa{{ Ba$_{1-x}$K$_x$Fe$_2$As$_2$ } }

\begin{document}

\title{Observation of superconducting vortices carrying a temperature-dependent fraction of the flux quantum}

\author
{Yusuke Iguchi$^{1,2}$, Ruby Shi$^{1,2,3}$, Kunihiro Kihou$^{4}$, Chul-Ho Lee$^{4}$, \\Vadim Grinenko$^{5}$, Egor Babaev$^{6}$, and Kathryn A. Moler$^{1,2,3,7}$}
\affiliation{$^{1}$Geballe Laboratory for Advanced Materials, Stanford University, Stanford, California 94305, USA\\
$^{2}$Stanford Institute for Materials and Energy Sciences, SLAC National Accelerator Laboratory, 2575 Sand Hill Road, Menlo Park, California 94025, USA\\
$^{3}$Department of Applied Physics, Stanford University, Stanford, California 94305, USA\\
$^{4}$National Institute of Advanced Industrial Science and Technology (AIST), Tsukuba, Japan\\
$^5$Tsung-Dao Lee Institute, Shanghai Jiao Tong University, Shanghai, China\\
$^6$Department of Physics, Royal Institute of Technology, SE-106 91 Stockholm, Sweden\\
$^{7}$Department of Applied Physics, Stanford University, Stanford, California 94305, USA
}

\baselineskip 24pt

\begin{abstract}
The magnetic response is a state-defining property of superconductors.
The magnetic flux penetrates type-II bulk superconductors by forming quantum vortices when the enclosed magnetic flux is equal to the magnetic flux quantum. The flux quantum is the universal quantity that depends only on the ratio of fundamental constants: the electron charge and the Planck constant. This work investigates the vortex state in the hole-overdoped \bkfa by using scanning superconducting quantum interference device (SQUID) magnetometry. We observed quantum vortices that carry only a fraction of the flux quantum, which vary continuously with temperature. This finding establishes the phenomenon that superconductors support quantum vortices with non-universally quantized magnetic flux. Furthermore, the demonstrations of the mobility of the fractional vortices and the manipulability of their positions open up a route for future fluxonics applications.
\end{abstract}

\maketitle

\section{Introduction}

A  classical electrically conducting fluid allows the creation of arbitrary vortices. By contrast, the fundamental property of an ordinary superconductor is the quantization of vorticity. This implies the quantization of the magnetic flux that these vortices induce. The magnetic flux $\Phi$ \cite{london1948problem} is defined as the integral of magnetic field ${\bf B}$ over the surface of the sample $\Phi=\int {\rm d}x {\rm d}y { \bf B}\cdot \hat{\bf n}$, where ${\bf B}$ is the magnetic field and $\hat{\bf n}$ is a unit vector normal to the surface. This quantization means that magnetic fields inside a superconductor can change only in discrete steps. Namely, an externally applied magnetic field penetrates a type-II superconductor, in increments $\Phi_0$ associated with the entrance of quantum vortices from the boundary of the sample \cite{Abrikosov1957}. The magnetic flux quantum $\Phi_0= hc/2e$ (in CGS unit) depends only on fundamental constants: the electron charge $e$, the speed of light $c$, and  the Planck constant $h$. The quantum vortices are mobile objects. Their stability is dictated by a non-trivial winding of the phase $\theta$ of the complex superconducting order parameter. The phase changes by $2\pi$ around the vortex core \cite{onsager1949statistical}. In type-I superconductors vortices cannot be induced by an externally applied magnetic field. On the other hand, a quenched type-I superconductor possesses stable vortices carrying an integer number of flux quanta. Under certain conditions, immobile objects carrying a half of the flux quantum were observed in an intrinsic Josephson junction formed on grain boundaries of $d$-wave superconductors \cite{kirtley1996direct} and on small rings of Sr$_2$RuO$_4$ \cite{jang2011observation} and $\beta$-Bi$_2$Pd \cite{Li2019Obs}.

In this paper we report scanning SQUID magnetometry on \bkfa ($x=0.77$). We show that, along with the standard vortices carrying single flux quantum, the material has vortex excitations that carry a fraction of the flux quantum. Importantly the fraction is not an integer but smoothly varies with temperature and hence the flux quantization is non-universal i.e. is not a function of only fundamental constants.

\section{Temperature dependent fractional flux in vortex}

To explore the magnetic properties of \bkfa, using the scanning SQUID susceptometer, we microscopically imaged the magnetic flux on the cleaved $ab$-plane  of single crystals (Figs. 1A,B) \cite{KihouJSPS2016,kirtleyrsi2016}. We conducted measurements  below the superconducting critical temperature $T_c $ (Fig. S1). By cooling  the sample in a small uniform magnetic field of 3.5 mG, we observed that below $T_c$, the system formed dilute configurations of conventional vortices that coexisted with isolated vortex-like objects carrying a small fraction of magnetic flux (Fig. S2). In order to determine whether these objects are vortices and to estimate the total flux trapped in the object, further measurements were performed. By cooling through $T_c$ with a small local magnetic field from the scanning SQUID field coil, we also observed the conventional vortices and the fractional vortex-like objects (Figs. 1C-E, Fig. S3). We observed both the conventional vortex and the fractional vortex-like objects appearing at the same location in different cooling cycles, where the sample was cooled down to 10.5 K from 25 K with fields (Figs. 1C,E). Hence the specific area in the sample did not determine the character of a flux-carrying object. To characterise further the nature of the object, a fractional antivortex object was created at the same area in different cooling cycles (Fig. 1D).

To obtain the simplest estimate for the magnetic field penetration depth $\lambda$, we simulated the magnetic field of the conventional single-quantum vortex with a point source magnetic monopole field with the total flux $\Phi_0$ \cite{Kirtleysst2016}. We roughly estimated the magnetic field penetration length, $\lambda$ from this point source model. We fitted it with the spacing between the sample surface and the pickup loop center, $z_0 =$ 1250 nm (Fig. 1H), yielding the magnetic field penetration length, $\lambda \approx 2.3$ $\mu$m. A fitting of the vortex object in Fig. 1C with $\lambda$ = 2.3 $\mu$m and the fractional point source, gave a very different result. In contrast to the conventional vortex results, we obtained the  fraction of the flux quantum  $\Phi_{\rm F} \approx 0.3\Phi_0$ at $T$ = 11.0 K in the vortex-like object (Fig. 1F). 

By measuring the magnetic flux at different temperature we observed that  the flux of this object was temperature-dependent. 
The magnetic field amplitude was also very different from that of a conventional vortex and half-quantum objects that form on grain boundaries of a $d$-wave superconductor \cite{Kirtley1999Temp}. We observed that the amplitude of the peak magnetic flux from the fractional vortex-like object decreased with decreasing temperature (Fig. 2A). This is in strong contrast to the conventional vortex. The usual vortex carries temperature-independent single flux quantum. Hence the peak amplitude of magnetic field increases with decreasing temperature because of the temperature dependence of the magnetic field penetration depth (Fig. 2A). To estimate the temperature dependence of the fractional flux $\Phi_{\rm F}$, we fitted the cross sections of the fractional vortex-like object at several temperatures as we show in Figs. 1F-G (Figs. 2B,C). We observed fractional vortex-like objects in different areas of the sample (Figs. S4,S5). Importantly, the obtained fraction $\Phi_{\rm F}/\Phi_0$ showed the similar temperature dependence at three tested regions (Fig. 3), pointing towards the fact that these objects are unconventionally quantized fractional vortices.

\section{Mobility of fractional vortex}
The fraction of the flux quantum gradually decreased with decreasing temperature, and below $T/T_c = 0.8$, it could not be resolved from the background noise. That can be due to that the fraction of carried flux further drops to very small value or/and the effect of magnetic flux delocalization of fractional vortices \cite{Babaev2009Mag}. When we cooled to 3 K, well below the temperature where we can resolve fractional vortex objects, then heated back above 9 K, we observed the re-emergence of the fractional vortex  objects with similar magnetic flux pinned at the same location but sometimes cooling and subsequent heating resulted in disappearance of the object from the scanned area.

To determine whether the observed vortex-like object is a true quantum vortex, a mobility of the object should be demonstrated. We checked that by monitoring the positions of these objects when we cooled or heated the sample.
We found that fractional vortex-like objects sometimes randomly moved to other positions (Figs. 4A,B). The observed process was akin to the mobility of conventional vortices moving between pinning positions. Next we were able to manually change the position of the fractional vortex-like objects by applying a local repulsive force by creating the field using the scanning SQUID field coil (Figs. 4C,D). We demonstrated that the scanning SQUID manipulated both the fractional vortex-like object and the conventional vortex at the same time (Fig. 4D). Note that in a very different system of strongly layered superconductors, using magnetic probe manipulation one can break a vortex  into two small magnetic excitations which are described as a broken stack  of pancake vortices \cite{Luan2009Magnetic}.
In contrast, we did not observe such behavior when manipulating integer flux vortices. Moreover some of the fractional vortex-like objects appeared without any other fractional vortex-like objects within 50 $\mu$m area (Figs. S2,S3).
These observed robustness and mobility established that the observed objects represented quantum vortices.

\section{Discussion}
 
The material \bkfa has the multiband electronic structure (see e.g. \cite{Maiti2013,Boeker2017}). Furthermore previous experiments obtained the evidence that \bkfa spontaneously breaks the time-reversal symmetry \cite{Grinenko2017,grinenko2020superconductivity,grinenko2021state}. The evidence of multiple broken symmetries dictates that there are several components of the order parameter. Hence, in principle, there are necessary degrees of freedom to support several types of vortices associated with phase windings in different components of the order parameter.

For a multiband superconductor with bands labeled by a band index $j$, $\psi_j=|\psi_j|e^{i\theta_j}$, in a simplest model the current will be given by a sum of contributions from different bands ${\bf J}=\sum_j({e \hbar}/{i2m})
(\psi_j^*\nabla\psi_j-\psi_j\nabla\psi^*_j) -({e^2 }/{mc})|\psi_j|^2 {\bf A}$, where ${\bf A}$ is the vector potential. Consider the case where  $\theta_1$ of the first band has a $2\pi$ winding: $\oint_\sigma \theta_1 {\rm d} l=2\pi $.
Then one can use the above expression to calculate the magnetic flux enclosed in the resulting vortex. This can be done by  integrating the vector potential over a path $\sigma$ located far away from vortex where $\bf J = 0$. That gives the magnetic flux of a vortex: $\Phi= \oint_\sigma {\bf A} {\rm d} l =\oint_\sigma  \nabla\theta_1 {\rm d} l({\hbar c}/{e}) {|\psi_1|^2}/{\sum_j |\psi_j|^2}=\Phi_0{|\psi_1|^2}/{\sum_j |\psi_j|^2}$. That quantity is no longer a function of just fundamental physical constants. Instead it depends on microscopic detail and temperature. This is because the ratio of the fields $|\psi_j|$ associated with superconducting components in different bands, have, in general, different temperature dependencies. Note  that in certain cases one cannot attribute different complex fields to different bands and multicomponent theory arises for fields which are associated with linear combinations of gaps in different bands \cite{Garaud2017}, but the argument  remains similar. 

There are factors that can  suppress, or in some cases prohibit the formation of fractional vortices in multiband superconductors. First, in superconductors the components are electrically charged. The unavoidable electromagnetic inter-component interaction tends to confine fractional vortices into integer flux vortices in the bulk of a multiband superconductor \cite{Babaev2002_vortices}. This is the principal difference compared with multicomponent superfluids where vortices carry no magnetic flux and no such electromagnetic confinement of vortices exists. Similarly the interband coupling tends to lock phases in different bands and hence to confine fractional vortices into one-flux-quanta composite objects. While, theoretically, pinned fractional vortices are allowed to form in a bulk of superconductor, these have not been previously observed in multiband superconductors such as MgB$_2$ and pnictides \cite{Nishio2010Scan,Li2011Low,Kalisky2011Behavior,Zhang2019Imaging}, nor in multiband materials where time-reversal symmetry breaking was reported \cite{Bjornsson2005Scanning,Kirtley2007Upper,Hicks2010Limits,Iguchi2021Local}.

As emphasized in the above, the existence of fractional vortices is, in principle, possible due to the multiband character of a material and does not require extra broken symmetries. Nonetheless we note that experiments indicate that \bkfa is a so-called $s+is$ superconductor \cite{Grinenko2017,grinenko2020superconductivity,grinenko2021state,Vadimov2018}. This state is defined as a state where the interband coupling is frustrated. The precise microscopic model for that state is not known. However, based on the calculations in the framework of minimal phenomenological Ginzburg-Landau model, it was suggested that under certain conditions the frustration promotes the formation of mobile fractional vortices in individual bands \cite{garaud2014domain}. Further theoretical studies are needed to connect the observation of fractional vortices with concrete microscopic models of \bkfa. 

Our observations demonstrate that vortices carrying temperature-dependent fraction of the flux quantum are possible in superconductors. Dynamics, control and manipulations of these objects are promising new directions that will give insights both into basic questions of superconductivity and into the nature of superconductivity in \bkfa. They also open a new research avenue of potentially using these objects in fluxonics-based cryogenic computing\cite{Miyahara1985Abrikosov,Golod2015Single}.

\section*{Acknowledgments}
The authors thank N. Nandi and Julien Garaud for fruitful discussions. This work was primarily supported by the Department of Energy, Office of Science, Basic Energy Sciences, Materials Sciences and Engineering Division, under Contract No. DE- AC02-76SF00515. E.B. was supported by the Swedish Research Council Grants No. 2016-06122, 2018-03659. V.G. was supported by DFG GR 4667/1 and the Würzburg-Dresden Cluster of Excellence on Complexity and Topology in Quantum Matter-ct.qmat (EXC 2147, Project ID 390858490).

\clearpage

\begin{figure*}[htb]
\begin{center}
\includegraphics*[width=16cm]{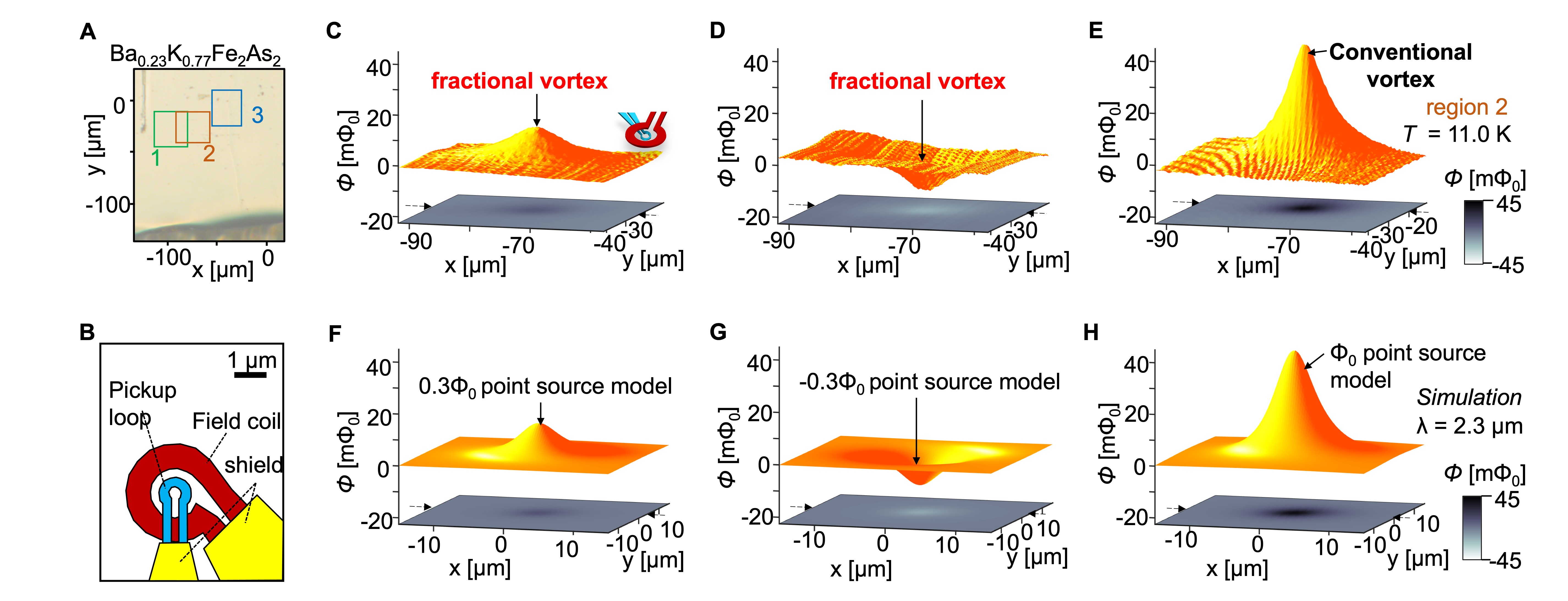}
\caption{{  Scanning SQUID imaging fractional vortices and conventional vortices at the same area of Ba$_{1-x}$K$_{x}$Fe$_2$As$_2$.} ({\bf A}), Optical image of the sample with scan regions 1,2 and 3. ({\bf B}), Pickup loop and field coil of the SQUID susceptometer are covered with superconducting shields except for the loop area to detect local magnetic flux. ({\bf C}-{\bf E}), Isolated ({\bf C}) fractional vortex carrying approximately 0.3 of the flux quantum, ({\bf D}) fractional antivortex, and ({\bf E}) conventional vortex appear in different cooling cycles. ({\bf F}-{\bf H}), Simulation of ({\bf F}) 0.3$\Phi_0$, ({\bf G}) -0.3$\Phi_0$, and ({\bf H}) $\Phi_0$ point source models show similar results of ({\bf D}) and ({\bf E}), respectively.}
\end{center}
\end{figure*}

\begin{figure*}[htb]
\begin{center}
\includegraphics*[width=16cm]{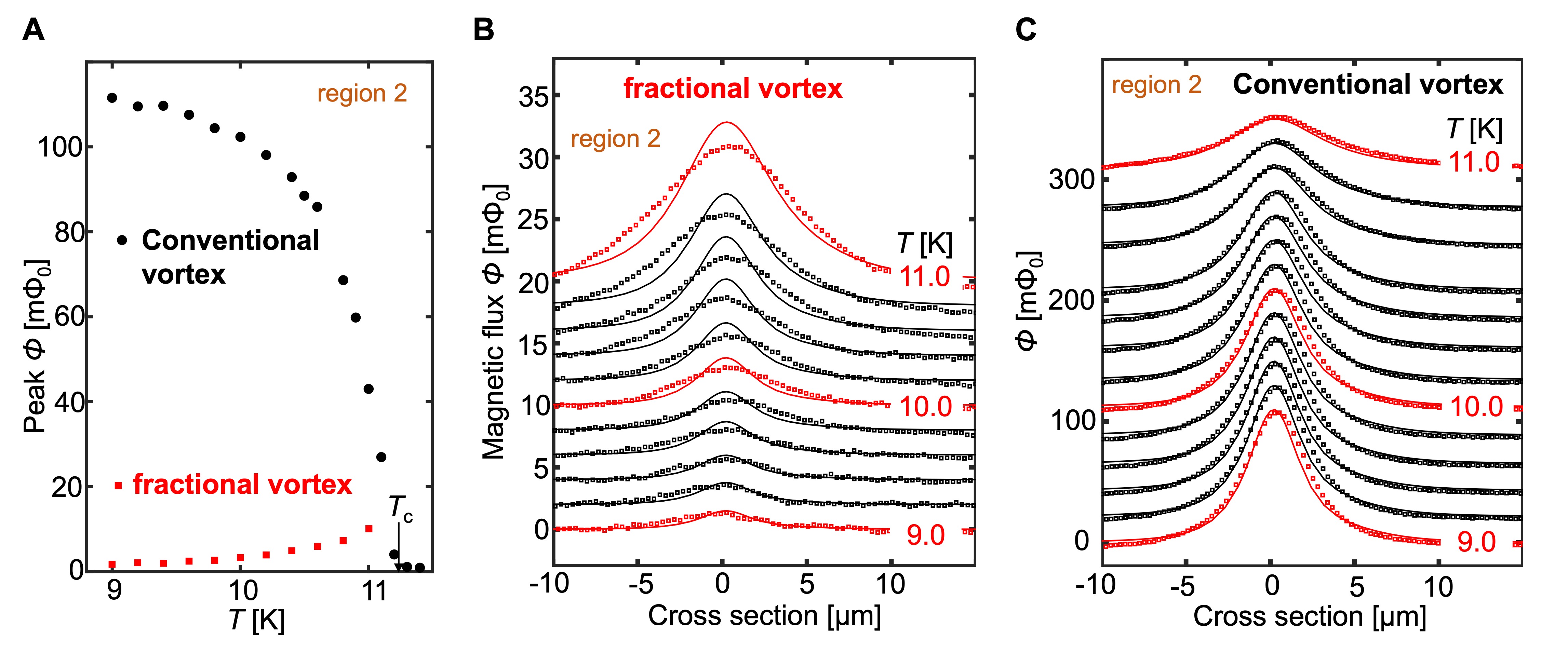}
\caption{{  Temperature dependence of the magnetic field of fractional vs conventional vortex.} ({\bf A}), Comparison of the temperature dependence of the maximum flux through the pickup loop with the loop above centers of the fractional vortex and the conventional vortex. The measurement indicates that the carried fraction of the flux quantum drops continuously as a function of temperature. ({\bf B}, {\bf C}), Cross sections of ({\bf B}) the fractional vortex and ({\bf C}) the conventional vortex along $x$ axis, where the $y$ position is indicated by dashed arrows in Figs. 1D,E. Solid lines are fitting results of point source models.}
\end{center}
\end{figure*}

\begin{figure*}[htb]
\begin{center}
\includegraphics*[width=7.5cm]{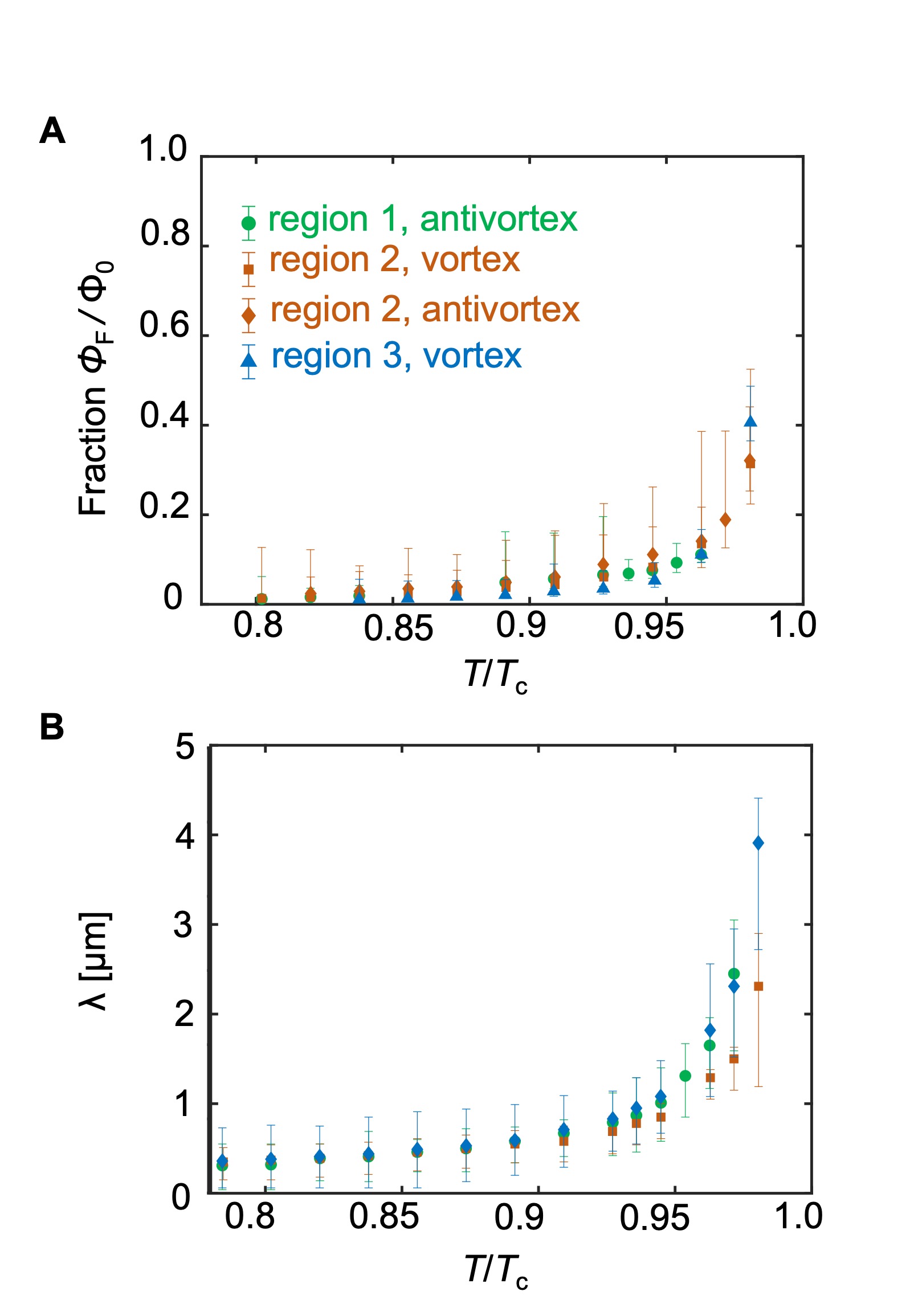}
\caption{{  Temperature dependent and ubiquitous magnetic flux in vortex.} ({\bf A}), The total magnetic flux in the fractional vortex $\Phi_{\rm F}$ at three regions obtained from fittings in Fig. 2B and Fig. S4 show the ubiquitous temperature dependence. ({\bf B}), Temperature dependence of the penetration depth obtained from the fittings in Fig. 2C and Fig. S5 are used to fit fractional vortices at same regions. The $T_c$ obtained from the scanning SQUID susceptometry (Fig. S1).}
\end{center}
\end{figure*}

\begin{figure*}[htb]
\begin{center}
\includegraphics*[width=13cm]{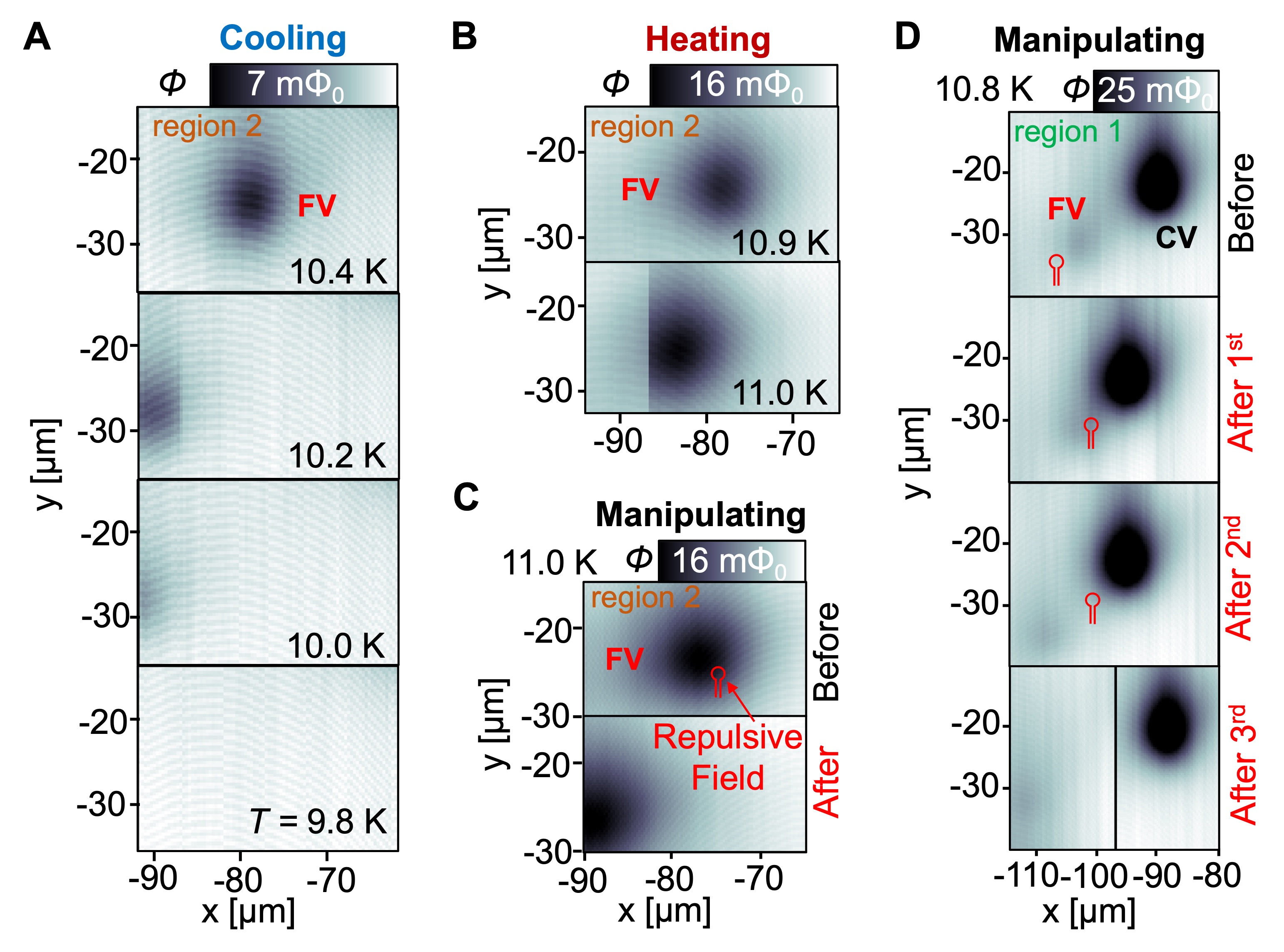}
\caption{{  Mobility of the fractional vortex.} ({\bf A}, {\bf B}), The fractional vortex (FV) moves while ({\bf A}) cooling and ({\bf B}) heating, which randomly happens. ({\bf C}, {\bf D}), Scanning SQUID susceptometer manipulates ({\bf C}) the FV and ({\bf D}) both of the FV and the conventional vortex (CV) by applying the repulsive local field. The schematic image of the field coil indicates the location where the local field is applied.}
\end{center}
\end{figure*}

\newpage
\clearpage

\setcounter{figure}{0}
\setcounter{equation}{0}
\renewcommand{\thefigure}{S\arabic{figure}}
\renewcommand{\theequation}{S\arabic{equation}}

\onecolumngrid
\appendix
	
\begin{center}
	\Large
	{Supplemental Material for \\\lq\lq Observation of superconducting vortices carrying a temperature-dependent fraction of the flux quantum \rq\rq} \\by Iguchi $et$ $al.$
\end{center}

\section*{Materials and Methods}
\subsection*{Sample}
All measurements were performed on the \bkfa sample with $x =$ 0.77 used in Ref.~({\it 17}) 
The doping level of K for the single crystals was determined using the relation between the $c$-axis lattice parameter and the K doping obtained in previous studies~({\it 7}) 
The sample was previously investigated by several different methods. The data are summarized in Figs.~2 and~3 from Ref.~({\it 17}) 
In particular, this sample shows a strong spontaneous Nernst signal indicative for broken time-reversal symmetry.
After the measurements, we cleaved the sample surface again for the scanning SQUID microscopy measurements.

\subsection*{Scanning SQUID Microscopy}
We used scanning SQUID microscopy to locally obtain the dc magnetic flux and ac susceptibility on the cleaved $ab$-plane of single crystals of Ba$_{0.23}$K$_{0.77}$Fe$_{2}$As$_{2}$ at temperatures varying from 3.0 K to 25 K using a Bluefors LD 4K refrigerator. Our scanning SQUID susceptometer had two pickup loop and field coil pairs configured with a gradiometric structure~({\it 8}) 
The inner radius of the pickup loop was 400 nm, and the distance between the pickup loop and the sample surface was $\sim$1200 nm when the scans were performed. The pickup loop provides the local dc magnetic flux $\Phi$ in units of the flux quantum $\Phi_0$. The pickup loop also detects the ac magnetic flux $\Phi^{ac}$ in response to the ac magnetic field $He^{i\omega t}$, which was produced by an ac current of $|I^{ac}| =$ 1 mA at frequency $\omega/2\pi=$ 1.223 kHz through the field coil, using an SR830 Lock-in-Amplifier. Here we report the local flux $\Phi$ and the local ac susceptibility as $\Phi^{ac}/|I^{ac}|$ in units of $\Phi_0/A$.

\subsection*{Simulation of point source magnetic field}
The numerical simulation of a point source magnetic monopole field $\Phi_{\rm sim}$ with total flux $\Phi=\Phi_0$ includes the SQUID structure~({\it 9,10}) 
In this model, the magnetic monopole is set at $z=0$, where the sample surface is at $z=\lambda$ and the center of pickup loop is at $z=z_0+\lambda$, where the magnetic monopole field is defined as $H(\vec r) = \Phi_0z/r^3$.
We simulated the total flux through the pickup loop area from the point source magnetic monopole. The penetration depth is obtained by fitting the experimental data with the simulated data. The error bars are assigned by changing the parameter of $\lambda$, finding least squares fits as a function of the other parameters, with a doubling of $\chi^2$ between fit and experimental values to obtain the penetration depth~({\it 10}) 
Next, to estimate the fractionally quantized flux $\Phi_{\rm F}$, we used $(\Phi_{\rm F}/\Phi_0)\Phi_{\rm sim}$ to fit the experimental results, where the $\Phi_{\rm sim}$ is obtained from the fitting of the conventional vortex at the same region at corresponding temperature because we assumed that the penetration depth was the same at the same region at the same temperature. The error bars are assigned by changing the parameter of $\Phi_{\rm F}$, finding least squares fits as a function of the other parameters, with a doubling of $\chi^2$ between fit and experimental values.

\clearpage

\begin{figure}[htb]
\begin{center}
\includegraphics*[width=7.5cm]{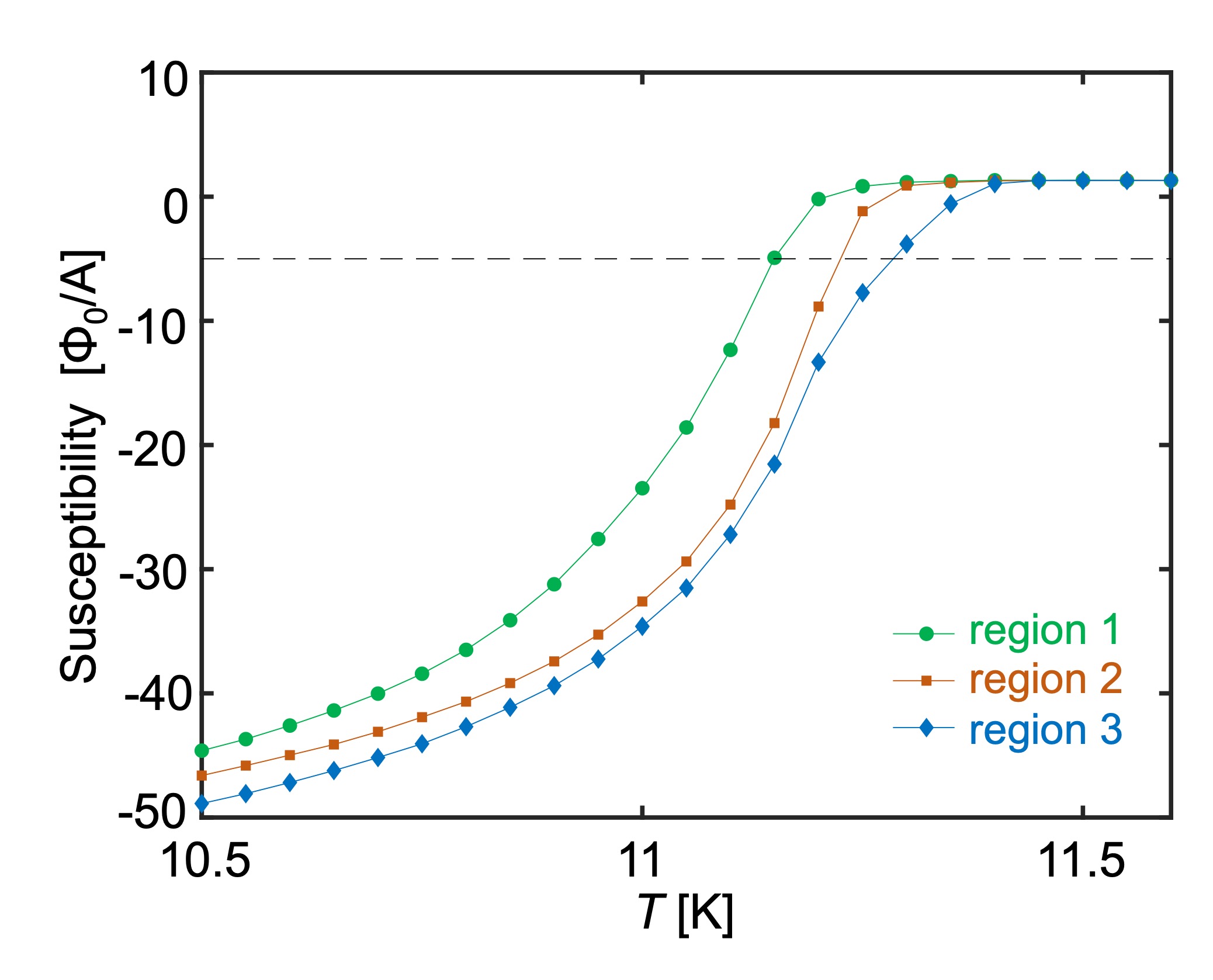}
\caption{ {   Scanning SQUID susceptometry at all regions.} $T_c$ used in Figs. 2,3 are obtained as the averaged value of temperatures where the susceptibility is -0.5 $\Phi_0$/A (dashed line).}
\end{center}
\end{figure}

\begin{figure*}[htb]
\begin{center}
\includegraphics*[width=15cm]{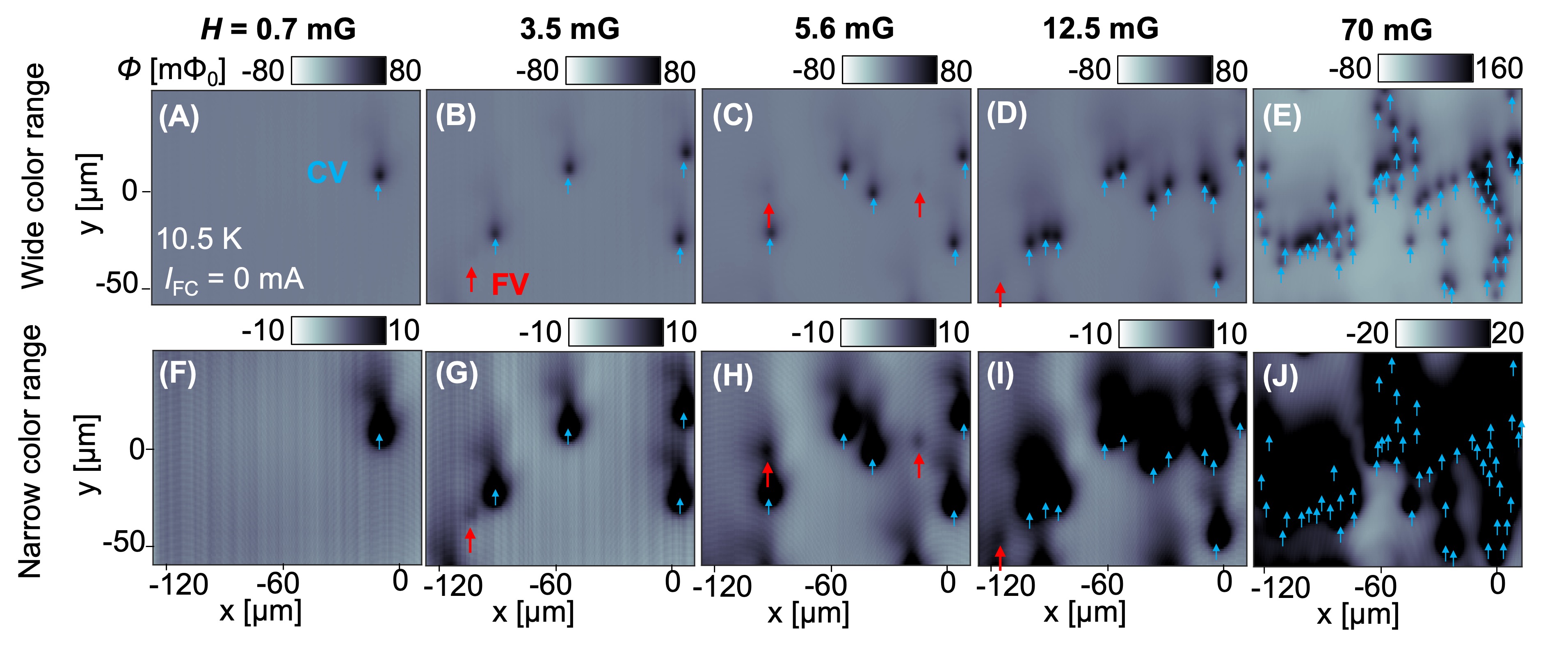}
\caption{ {  Fractional vortices are visible in scanning SQUID magnetometry at small uniform magnetic fields.} Scanning SQUID magnetometry at the large area including region 1,2, and 3 after uniform field $H$ cooling from 25 K to 10.5 K. (A-E) The magnetometry scans are plotted in wide color ranges to see the conventional vortices (CVs). (F-J) The same magnetometry scans with (A-E) are plotted in narrow color ranges to see the fractional vortices (FVs). The CVs are highlighted with blue arrows and the FVs are highlighted with red arrows. There are "tails" of vortex fields that are due to flux penetration through and around the superconducting shield layer in the area of the pickup loop~({\it 8}), 
thus in areas very close to a CV it is difficult to distinguish a FV from an artificial tail.}
\end{center}
\end{figure*}

\begin{figure*}[htb]
\begin{center}
\includegraphics*[width=15cm]{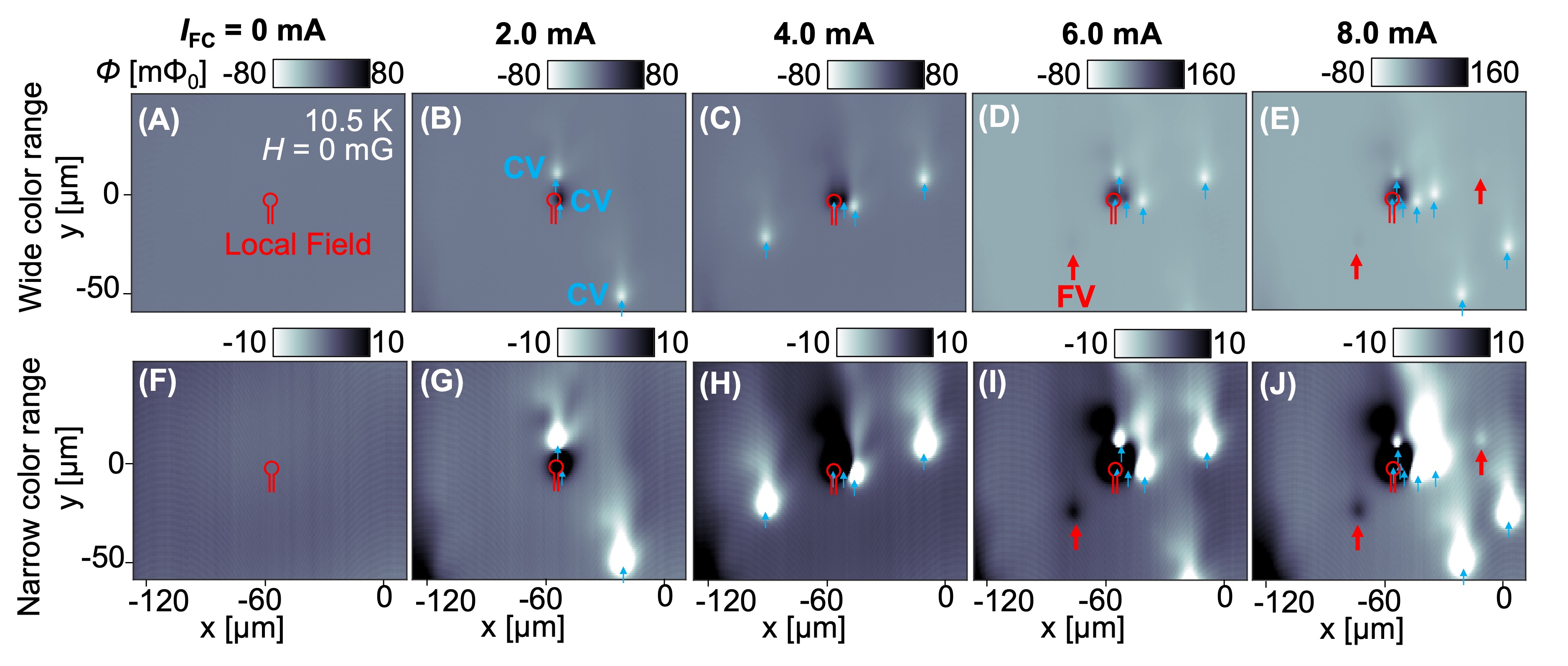}
\caption{ {\bf Fractional vortices are induced by local magnetic fields cooling.} Scanning SQUID magnetometry at the large area including region 1,2, and 3 after local field cooling from 25 K to 10.5 K. The local field is created by applying an dc current $I_{\rm FC}$ through the scanning SQUID field coil. (A-E) The magnetometry scans are plotted in wide color ranges to see the conventional vortices (CVs). (F-J) The same magnetometry scans with (A-E) are plotted in the narrow color range to see the fractional vortices (FVs). The CVs are highlighted with blue arrows and the FVs are highlighted with red arrows. The schematic image of the field coil indicates the location where the local field is applied. }
\end{center}
\end{figure*}

\begin{figure*}[htb]
\begin{center}
\includegraphics*[width=15cm]{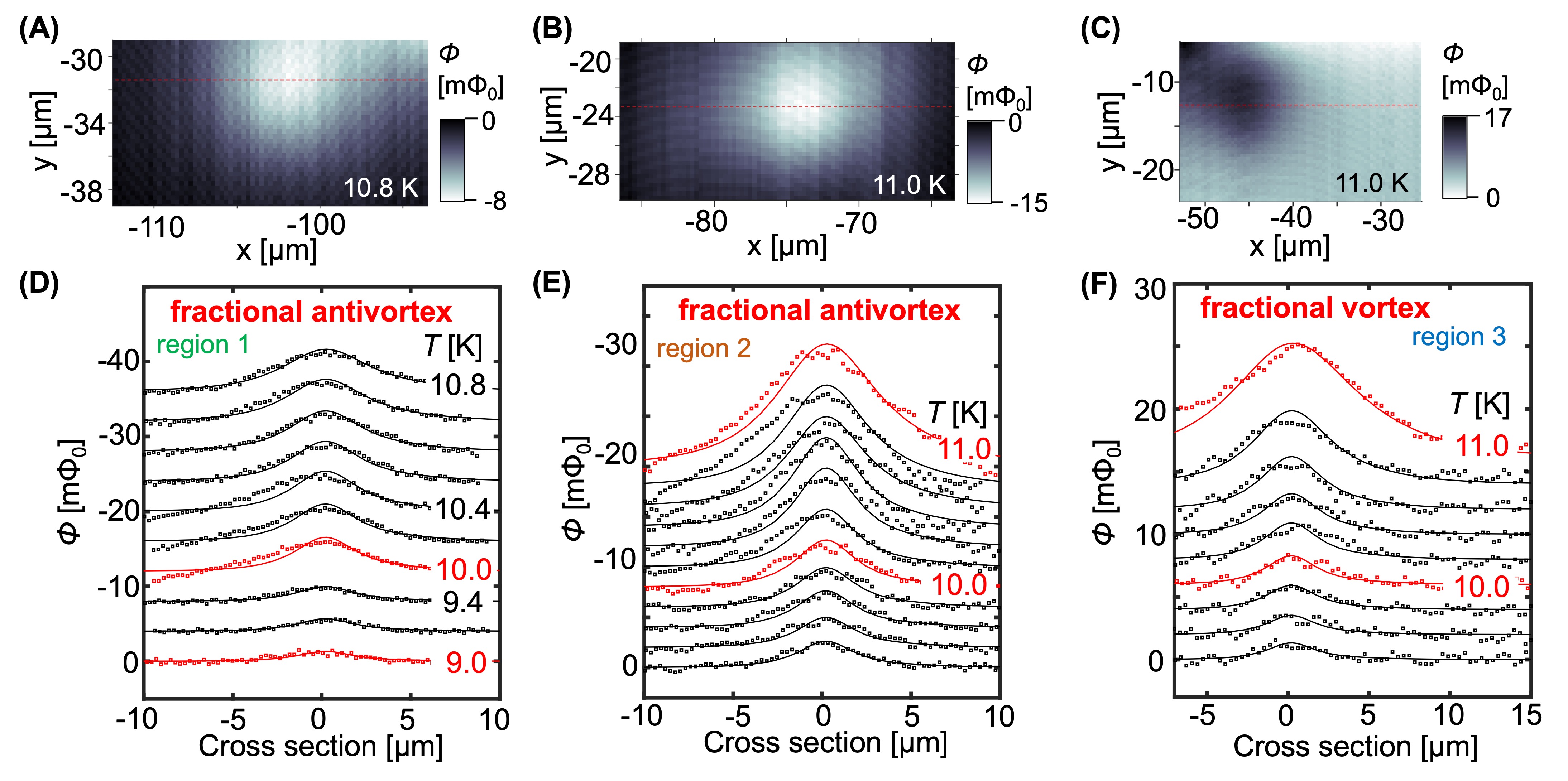}
\caption{{  Scanning SQUID magnetometry of fractional vortices at all regions.} (A-C), Scanning SQUID magnetometry of fractional vortices at region 1,2 and 3, respectively. (D-F), Cross sections of the fractional vortices at region 1,2, and 3 with the fractional flux quantum point source model fitting curves.}
\end{center}
\end{figure*}

\begin{figure*}[htb]
\begin{center}
\includegraphics*[width=11cm]{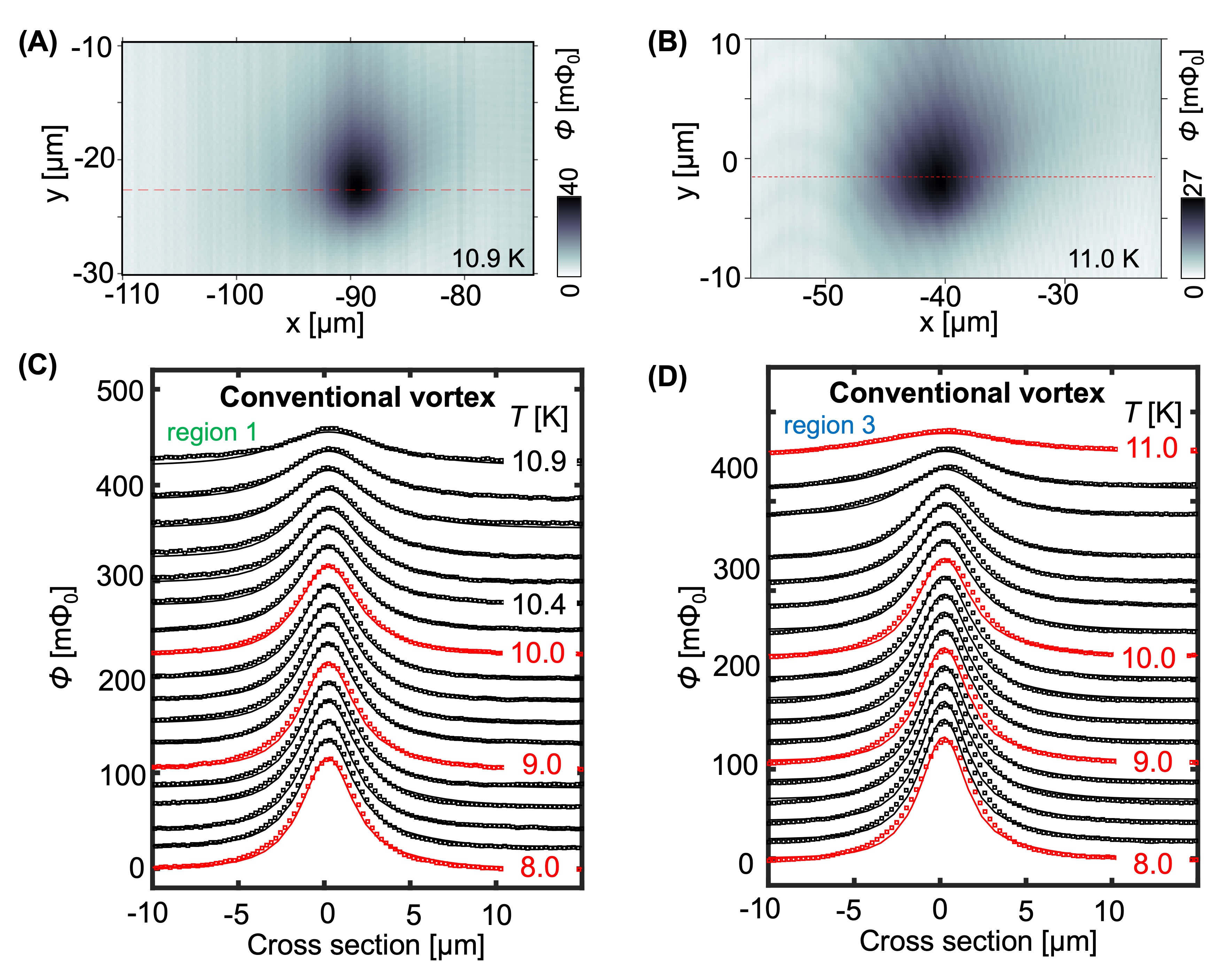}
\caption{{  Scanning SQUID magnetometry of conventional vortices at region 1 and 3.} (A,B), Scanning SQUID magnetometry of the conventional vortices at region 1 and 3. (C,D), Cross sections of the conventional vortices at region 1 and 3 with the flux quantum point source model fitting curves.}
\end{center}
\end{figure*}

\end{document}